\documentclass{article}

\usepackage{arxiv}

\usepackage[utf8]{inputenc} 
\usepackage[T1]{fontenc}   
\usepackage{hyperref}      
\usepackage{url}            
\usepackage{booktabs}      
\usepackage{amsfonts}       
\usepackage{nicefrac}       
\usepackage{microtype}      
\usepackage{graphicx}
\usepackage{subcaption}
\usepackage{textcomp}
\graphicspath{ {./images/} }

\title{Less is more: AI Decision-Making using Dynamic Deep Neural Networks for Short-Term Stock Index Prediction}

\author{
CJ~Finnegan\thanks{corresponding author }  \\
        Plotinus Asset Management \\
	Camana Bay \\
	Grand Cayman KY1-1204 \\
	Cayman Islands \\
	\texttt{cj.finnegan@plotinus.tech} \\
	\And
	James F.~McCann \\
	 Plotinus Asset Management \\
	Camana Bay \\
	Grand Cayman KY1-1204\\
	Cayman Islands \\
	\And
	Salissou Moutari \\
	School of Mathematics and Physics\\
	Queen's University Belfast\\
	Belfast BT7 1NN\\
	Northern Ireland, UK
 }

\begin{document}
\maketitle
\begin{abstract}
In this paper we introduce a  multi-agent deep-learning method  which trades in the Futures markets based on the US S\&P 500 index. The  method (referred to as Model A) is an innovation founded on existing well-established machine-learning models which sample market prices and associated derivatives in order to decide whether the investment should be long/short or closed (zero exposure), on a day-to-day decision. We compare the  predictions with some  conventional machine-learning methods namely, Long Short-Term Memory, Random Forest and Gradient-Boosted-Trees.
Results are benchmarked against a {\em passive model}  in which the Futures contracts are held (long) continuously 
with the same exposure (level of investment). Historical tests are based on daily daytime trading carried out over a period of 6 calendar years (2018-23). We find that Model A outperforms the passive investment in key performance metrics, placing it
within the top quartile performance of US  Large Cap active fund managers. Model A also outperforms the  three machine-learning classification comparators over this period. We observe that Model A is extremely efficient (doing less and getting more) with an exposure to the market of only 41.95\% compared to the 100\% market exposure of the passive investment, and thus provides increased profitability with reduced risk. 

\begin{center}
\small{\textsf{\textbf{Performance Metrics (Based on Monthly Returns Jan 2018 – Dec 2023).}}}

\vspace{0.25cm}

\begin{tabular}{lrr} \toprule
{ }  &  {\sf  Passive } & {\sf  Model A } \\ \midrule
{\sf  Annualized Return }  & 3.09\% & 14.92\% \\
{\sf   Annualized Volatility }  & 13.10\% & 10.26\% \\
{\sf    Alpha (Annualized)  }  & 0.00\% & 11.60\% \\ \midrule
{\sf    Beta (Monthly)  }  & 1.00\ \ \ \ & 0.18\ \ \ \  \ \\
{\sf     Sharpe Ratio   }  & 0.15\ \ \ \  & 1.16\ \ \ \  \ \\
{\sf    Sortino Ratio } & 0.22\ \ \ \  &  2.97\ \ \ \  \ \\  \bottomrule
\end{tabular}
\end{center}

\end{abstract}

\section{Introduction}

 Applications of machine learning to trading has been an active area of research \cite{hen19,jia19,kum22} and the mountain of  reports published (and unpublished) are witness to its importance \cite{han21}. 
The challenge is predicting future prices, thus informing which instruments to buy and sell and in what volumes, over a given time horizon. Such a problem should be well-suited to machine-learning \cite{lim21}, especially considering the learning goals are very clearly defined. However a consistent and robust 
approach has been elusive.  Given the plethora of financial variables, the abundance of 
model parameters, the randomness of fluctuations and the non-stationarity of the variables, the difficulties are obvious. There are practical considerations as well, some  academically-successful  models often overlook the friction of `spreads' and costs of trading that rapidly erode profit in the real world.

An  accepted measure of excellence for active investment funds is  consistently outperforming  established benchmarks  measured in higher net profit and lower risk. 
Typically a benchmark is a  passive strategy in which funds are allocated in advance (empirically) among a range of  investments and not reallocated thereafter.  Specifically  these passive investments 
are `long'  positions  in company stocks  in the anticipation their values will increase in the future. 
The most   well-known (and largest) financial market in equities is the US stock exchange, and the flagship 
benchmark  is the S\&P 500  index which consolidates the largest 500+ companies weighted  by market capitalisation. A naive passive investment would be to 'track' this index through a  low-cost exchange-traded fund. 
 
Based on data, year ending December 31, 2023, over a five-year period 78.68\% of US Large-Cap funds underperform the S\&P 500 index \footnote{S\&P Global, (2023) SPIVA® Scorecard, S\&P Dow Jones Indices}. Any strategy which consistently  outperforms 
the (naive and passive) S\&P 500 index would be within the top quartile of active funds. 
 In this study we have chosen the S\&P 500 index, or more precisely the futures contracts derived from the index, as the single tradable 
 instrument.  The corresponding  benchmark is the passive (long only) investment  in S\&P 500  futures.

\section{Data and Methods}
\label{sec:market}

\begin{figure}[htbp]
	\centering
	 \caption{The Mediocre Daytime Contribution to the Growth of the S\&P 500 
	Futures Over Twenty Years (Jan 2004 – Dec 2023).}
	\vspace{0.25cm}
        \includegraphics[width=14cm]{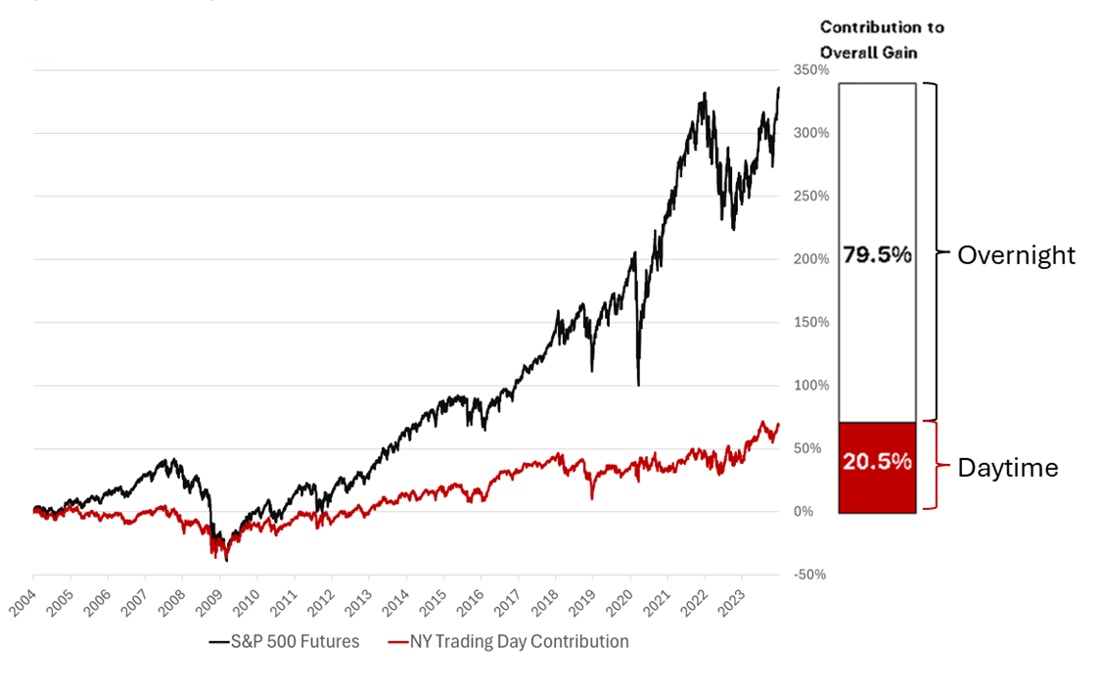}
	\label{fig:fig1}
\end{figure}

The Wall Street trading day from its opening to closing bell (09:30 – 16:00 ET) represents both psychologically and in trading volumes the most actively traded portion of the daily S\&P 500 futures market, which trades for 23 hours each day. In spite of this fact, over the last twenty years 
(Figure~\ref{fig:fig1}) the New York trading day only contributes 20.5\% of the overall gain in the S\&P 500 futures.

The poor performance of the passive daytime trade, suggests  an active method over this period could improve yields. In contrast to the passive, which is always 'long'  at the daytime open, the machine learning action would have the option of going long, short or closed. Typically a {\em  big data}  approach  aims at finding a  synthetic key indicator within the set of features. Computational resources aren't a limiting factor, and  vast amounts of data can be used and digested, in principle. However, it is well-known that a surfeit of data can be a curse as much as a blessing. The S\&P 500 is a challenging example in that it is composed of 500 elements moving in unpredictable correlations. Given such diverse underlying elements, the high-dimensionality and randomness of variables, we asked the question, ‘is there empirical evidence to justify the use of more data?’ 

The fund management industry first and foremost is responsible to investors to provide them with a long-term investment process, a consistent means to provide better diversified risk-adjusted returns. Active managers are employed to use their skills to outperform sector benchmarks. 
We determined that there is no way to future-proof big data use, considering the variables/features required and relevance. Such data is vulnerable to potentially being deemed inoperable due to many different factors, regulation, monopolisation, deprecation to name a few and in the case of the use of alternative datasets, an inability to provide historical validation. Thus, we conclude that big data methods, though possibly being able to offer useful trading signal solutions, have difficulty responding to the investors' pre-eminent requirement, to have a consistent, long-term low-risk  process. We begin the discussion with various methods 
that can be applied to the problem, before introducing our Model A approach which  illustrates how a ‘small data’ approach can work in practice.

In keeping with the objective to test the viability of trade decision-making models using curtailed data inputs, data was restricted to standard exchange-based data. Only CME E-mini S\&P 500 futures contract (ticker symbol ES) price and volume data and the CBOE volatility index (ticker symbol VIX) price data from January 2018 to December 2023. Separately for the purposes of calculating performance metrics,
 requiring the risk-free rate, Federal Reserve data for the Three-Month US Treasury Bill was used from January 2018 to December 2023.

\subsection{Long Short-Term Memory (LSTM) }

Deep Neural Networks are well-adapted to processing sequential data, and in particular time-series prediction \cite{lim21}. 
Each family of method  \cite{han21} has particular strengths and there is no general purpose best-in-class.
While Support-Vector Machines (SVM) classification methods have had success in this field, a consistently accurate technique is  the Long Short-Term Memory  (LSTM) network \cite{hoc97,yu20}. Even within a noisy data environment such as financial markets, 
this algorithm has been found to provide market pattern recognition in a rapid and accurate manner \cite{fis18}.

As with any model, the key step is selecting suitable input data variables. The architecture requires input on the number of layers, depth, and nodes per layer. In additional one can vary the sample size, the batch size, the sequence length, the error criterion, and the number of cycles (epochs) and so on. Given the large number of degrees of freedom mentioned, tuning such a network is not straightforward. 

In this context, we present a simple version of LSTM in order to illustrate the effectiveness of the model for binary classification to trade futures contracts. The model needs to instruct us on the correct open position in the S\&P futures market, beginning at the daily opening time and closing the position at the end of the day. If the futures price rises between the start and end of the day, that is the Open-to-Close price difference is positive, then a long position would be profitable and the correct classification. Thus, our goal is to train the network to anticipate price rises (falls) and thus to open a long (short) position, appropriately on each trading day.

After some experimentation and permutation, we selected 3 input variables: 
\begin{itemize}
\item[(a)]
 The CME E-mini S\&P 500 futures contract price at  daytime open (ticker symbol ES), on day $ t $.
\item[(b)] The CBOE volatility index (ticker symbol VIX) open price on day, $ t $.
\item[(c)]  The volume of futures contracts traded on the previous day, $ t-1 $.
\end{itemize}
Empirically, we settled on a sequence length of 20 (days) for each variable. 
Choosing variable (a) allows trend-following to be represented. The degree of randomness in this price means that auto-regression is of limited value. On the other hand, the market volatility, which is measured by variable (b) is very important for short-term forecasting, as is variable (c) which reflects the activity of the market, and these are often mirrored in price fluctuations.  The response variable is  binary, and transformed from the daytime price rise/fall: with $+1$ being a profit and $-1$ being a loss. Accordingly,  the network was used to take either a long
or short position each trading day.

The architecture is quite simple and thus convergence, with respect to epoch number, was rapid. We used a sample size of 250 days, and then ran the (closed-loop) prediction forward for a 50-day window. The LSTM was 
then retrained over 250 days before running for the next 50-day window. 
The 50-day testing window results were concatenated to create the long-run trading signal. 
The models converged during training and the method was profitable compared to the passive approach, 
although it produced fluctuation (volatility) values higher than the passive investment. Details are presented in the Results section. 

\subsection{Gradient-Boosted Trees and Random Forests}

The availability of cloud computing power along with data proliferation and widely accessible algorithms has fuelled machine learning classification models. Attempts to predict intrinsically unpredictable systems such as whether the market goes up or down from one day to the next, remain unsolved. Among the standard (and generic) classification models, decision trees using bagging and boosting have had limited success \cite{han21}. 

In particular, the ensemble approaches seem to avoid the pitfalls of overfitting while maintaining accuracy. The twin algorithms of Gradient Boosted Trees (GBT) and Random Forest (RF) methods have established themselves as the baseline for classification. 

Gradient Boosting is an iterative machine learning technique used to build a strong classifier from a set of weak classifiers. It builds an ensemble of models, typically decision trees, in a sequential manner.  Each new model is trained to correct the errors of the previous models by focusing on the residual errors. This is done by minimizing a specified loss function through gradient descent.  The final model is a weighted sum of all the individual models, and this iterative approach results in a strong predictive model that combines the strengths of many weak models \cite{fri01}. Gradient Boosting is known for its high accuracy and effectiveness, especially with structured data, but it can be computationally intensive and requires careful tuning to prevent overfitting \cite{che16,has09}.

Random Forest is an ensemble learning algorithm widely used for classification tasks. In contrast with Gradient Boosting, it constructs multiple decision trees during training and merges their outputs to improve accuracy and control overfitting. Each tree in the forest is built from a random subset of the training data, and each node in a tree is split based on a random subset of features. This randomness helps create a diverse set of trees, reducing the variance of the model and improving its generalization ability. The final class prediction is made by taking a majority vote \cite{bre01}. 

Random Forests are known for their robustness, ease of use, and ability to handle large datasets with high dimensionality. However, Random Forest can be computationally intensive and less interpretable compared to simpler models \cite{lia02,has09}.
However, when applied to the stock market the success of GBT and RF has been mixed \cite{yun21,nti20,vij20,zho19}. 
We use one Gradient Boosted Tree model and one Random Forest model. 

These decision-tree models aim at a prediction/classification of two categories (a) a positive daytime profit, and  (b) a  negative daytime profit. 
Our aim with each is simply to predict the daily change in price of the S\&P 500 index during daytime trading, or more specifically the futures contracts derived from the index, which is tradable on exchanges. 
If one predicts a price rise during the day, from market open to close, then it makes sense to take a ‘long’ position which is then closed at the end of the day. Conversely if the prediction (classification) is a price decrease, then a ‘short’ position is the decision. For our features/inputs we use futures prices from (a) CME E-mini S\&P 500 futures contract (ticker symbol ES) price data (open, high, low, close \& volume traded) and (b) the CBOE volatility index (ticker symbol VIX) price data (open, high, low, close).  

\begin{figure}[ht]
	\centering
	\caption{Schematic Representation of Model A.}
	\vspace{0.25cm}
        \includegraphics[width=12cm]{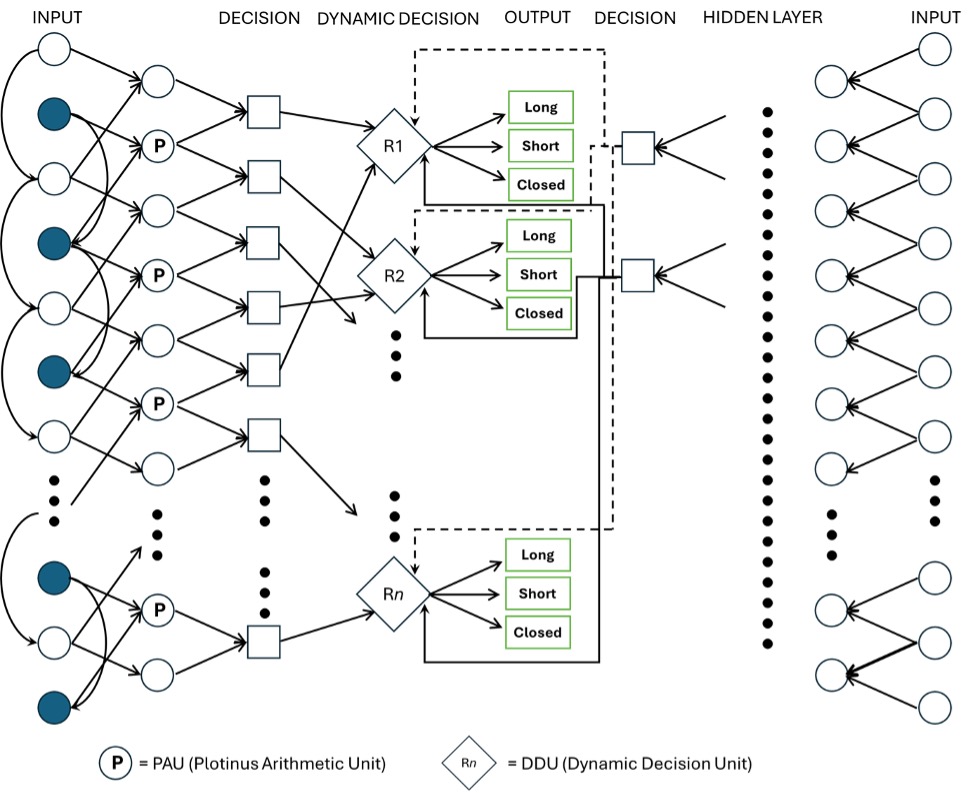}
 	\label{fig:fig2}
\end{figure}

\subsection{Dynamic Deep Neural Network – Model A} 

Our concerns around the long-term viability of using big data helped provide the criteria used to design Model A. It was built to use limited amounts of standard financial data with reinforcement learning applied \cite{zha21}. The method employs two agents acting autonomously but within the same data environment,  each tasked with optimizing its own short-term return by interacting with the environment and  the partner  agent.  In the version of the model we present, and the one being used currently by Plotinus Asset Management to trade in this market, the data environment was limited to: (a) CME E-mini S\&P 500 futures contract (ticker symbol ES) price data (open, high, low, close) and (b) the CBOE volatility index (ticker symbol VIX) price data (open, high, low, close). 
The architecture of the Model A is a multi-agent system, created from a hybrid of a deep neural network and decision trees. In acknowledging market flux, we wanted the model to be able to move beyond the constraints of the time-series data input it was provided with and develop a contextual understanding of market conditions, rather than time-dependent ones (such as trend or momentum). In order to do this, we enabled the model to firstly develop a decision-making context and then to be able to continuously reassess it: a process of reinforcement learning. This contextual reassessment was performed once every 24 hours as the objective was to successfully ascertain the direction of the market each day, over the duration of daytime trading. We used a sample size of 250 days and ran a (closed-loop) prediction forward for a single day window. Simultaneously, the decision-tree agent used the same input variables with a smaller sample size of features but using a sequence of 20 days, to provide the model with an curtailed assessment of near-term market behavior (approximately one trading month) to assist the decision-making process.

The model creates a {\em decision}  (long, short or closed) and supplements this by an action : the scale of investment to make. For example, if the given decision is long/short, the action can be zero, and thus effectively the long/short is corrected to a closed position. Thus, Model A has in effect three categorical outputs: long, short and closed. Conversely the LSTM, GBT and RF models do not have the closed output nor the additional corrective action and thus are pure binary decisions with the same exposure. In practice, this agent-based scaling action serves to suppress exposure, particularly in volatile markets. On the other hand, the scaling (exposure) can be amplified when the agent deems it prudent and profitable. So long as the decisions are accurate, when exposure is suppressed in such a manner the risk/reward ratio is reduced. In the results presented, the scaling used was either $+1$ for long, $-1$ for short and $0$ for a closed position.

\section{Results}

We firstly assess the classification performance using conventional accuracy metrics associated with a 
confusion matrix. In the following discussion 
an  {\em observed positive} is designated  
as a daytime positive profit, while a loss is labelled as an {\em observed negative}. Thus the {\em true positives} 
are positive (observed) profits which were predicted correctly.  The testing period was 
from the beginning of January  2018  to the end of December  2023, a total of six years, leading to $N=1509 $ observations. 

\begin{table}[htbp]
\small
	\caption{Prediction Accuracy.}
	\vspace{0.25cm}
	\centering
	\sffamily
	\begin{tabular}{lccccc} 
 &   &   &   { \textsf{\textbf{ Gradient }} }  & { \textsf{\textbf{  Random }}}  &         \\	
  &  { \textsf{\textbf{ Passive}} }   &   { \textsf{\textbf{ LSTM }}}  &  { \textsf{\textbf{ Boosted } }} & { \textsf{\textbf{  Forest  } }} &   { \textsf{\textbf{ Model A  }}}      \\	
   &  &   &   { \textsf{\textbf{ Trees }} } &    &       \\	
      &  &   &    &    &       \\	
Accuracy   & 54.61\% & 	50.23\% &	 51.76\%	& 50.89\%. & 	56.87\%    \\
Positive Predictive Value   & 54.61\%  & 54.65\%	 &  52.26\%  & 	50.55\% & 58.94\% \\
Negative Predictive Value     & - &   45.16\% & 51.08\% & 51.40\% & 52.97\% \\
\end{tabular}
	\label{tab:tab1}
\end{table}

Table~\ref{tab:tab1} presents accuracy/precision results of classification.  The first row (accuracy) is  the sum of true positive and true negatives divided by the   total number of observations, {\small ACC = (TP+TN)/N}. While the {\em positive predictive value}  is the ratio of true positives to total predicted positives  {\small PPV = TP/PP} and  {\em negative 
predictive value}  is the ratio of true negatives to total predicted negatives {\small NPV = TN/PN} \cite{luq19}. 

\begin{table}[h]
\small
	\caption{Summary Statistics (Based on Daily Daytime Returns Jan 2018 – Dec 2023).}
	\vspace{0.25cm}	
	\centering
	\sffamily
	\begin{tabular}{lccccc} 
 &   &   &   { \textsf{\textbf{ Gradient }}}   & { \textsf{\textbf{  Random }} } &         \\	
  &  { \textsf{\textbf{ Passive}} }   &   { \textsf{\textbf{ LSTM }}}  &  { \textsf{\textbf{ Boosted } }} & { \textsf{\textbf{  Forest   }}} &   { \textsf{\textbf{ Model A  }}}      \\	
   &  &   &   { \textsf{\textbf{ Trees }}}  &    &       \\	
      &  &   &    &    &       \\	
Mean &\ 0.02\%& \	0.03\%&	\ 0.03\%&\ 	0.03\%&\	0.05\% \\
Standard deviation &\ 0.96\%&\ 	0.96\%&\	0.96\%&\ 	0.96\%	&\ 0.56\%\\
Minimum &-5.64\%	&-3.68\%&	-5.64\%&	-4.65\%&	-2.92\%\\
Maximum &\ 4.80\%	&\ 5.64\%&\	4.65\%&\	5.64\%&	\ 4.97\%\\
Skew &-0.27\ \ \ \ 	&\ 0.61\ \ \ \ 	&-0.17\ \ \ \  &\	0.25\ \ \ \  &\	1.51\ \ \ \  \\
Kurtosis &\ 4.01\ \ \ \ 	&\ 3.91\ \ \ \ 	&\ 4.01\ \ \ \  &\	3.97\ \ \ \  &	15.35\ \ \ \  \\
\end{tabular}
	\label{tab:tab2}
\end{table}

In  the first row in Table~\ref{tab:tab1}, the {\em accuracy} for the passive (54.61\%)  simply refers to that fact that  824 out of 1509 days had a positive daytime return, showing that the positives and negatives are well-balanced in this sample. The low accuracy values across the models is testament to the difficulty of this prediction problem.  Regarding the negative predictions, the  machine learning models are quite weak. This should however be considered in the context of the accuracy of a passive short benchmark which in this sample is 45.39 \%.

Given the reward (daily profit) is itself stochastic, the profitability doesn't follow accuracy. Table~\ref{tab:tab2}, in which the descriptive statistics of the daytime returns (fractional profit) are presented, illustrates this. For instance, the passive has higher overall accuracy than the three classification models yet it is less profitable, as observed in its mean daytime return. Model A, although the best of the models at negative prediction is much better in true positives. Model A has an additional agent which implements the action/exposure, and this has a major impact on profitability and variability in the net returns. The sample standard deviations are uniform apart from the Model A result which involves the agent action reducing the fluctuations.

Table~\ref{tab:tab3}  provides a summary of typical standard performance metrics, defined in  \cite{bod21}, for example. While there are large maximum daily returns for the LSTM and  RF models accruing from short positions, the level of dispersion is rather high, and leads to a low Sharpe ratio (Table~\ref{tab:tab3}). It is worth noting that the Beta-values for the machine-learning approaches, which is a measure of the correlation between daily classifications and the market (passive)  is very low in all cases for the models. 
Perhaps the most salient feature is the Alpha which is the uncorrelated return which is significantly better for Model A.  The advantage of having a low Beta is that the investment is decoupled from the market lows and highs, which 
leads to a lower risk profile. Finally, in terms of the annualized return (Table~\ref{tab:tab3}) the results of Model A are very promising and would support development of this approach further. 

Model A shows a significant and consistent outperformance over the passive benchmark, with lower volatility. The second-best performer is the LSTM but its higher return than the passive comes at the expense of higher volatility and much deeper drawdowns (see Table~\ref{tab:tab4}). As mentioned before, the value of LSTM could be ascribed to the randomness of returns given the shortcomings in predictive accuracy.

\begin{table}[bp]
\small
	\caption{Standard Performance Metrics (Based on Monthly Returns Jan 2018 – Dec 2023).}
	\vspace{0.25cm}
	\centering
	\sffamily
	\begin{tabular}{lccccc} 
 &   &   &   { \textsf{\textbf{ Gradient }}}   & { \textsf{\textbf{  Random }}}  &         \\	
  &  { \textsf{\textbf{ Passive}}}    &   { \textsf{\textbf{ LSTM }}}  &  { \textsf{\textbf{ Boosted  }}} & { \textsf{\textbf{  Forest  } }} &   { \textsf{\textbf{ Model A  }}}      \\	
   &  &   &   { \textsf{\textbf{ Trees }} } &    &       \\	
      &  &   &    &    &       \\	
Alpha (annualised) &  \    0.00\%&\	6.88\%	&\ 4.00\%&\	4.27\%	&\ 11.60\% \\
Annualised Return &\ 3.09\%&	12.26\%&\ 	6.93\%&\	6.14\%&\	14.92\% \\
Average Return (Monthly) &\ 0.33\%&\	0.75\%&\	0.51\%&\	0.52\%&\	1.15\% \\
Average Gain (Monthly) &\ 2.79\%&\ 	3.24\%&\	2.87\%&\	3.43\%&\	2.45\% \\
Average Loss (Monthly) &-3.13\%&	-2.74\%&	-1.99\%&	-2.24\%	&-1.44\% \\
Annualized Volatility &\ 13.10\%&	17.46\%&	11.30\%&	13.86\%& 	10.26\% \\
Beta &\ \ \ 1.00\ \ \ \  & \ \	0.09\ \ \ \ &\	0.12\ \ \ \ &\ \	0.02\ \ \ \  &\ \	0.18\ \ \ \  \\
Sharpe Ratio &\ \ \ 0.15\ \ \ \  &\ \	0.40\ \ \ \ 	&\ 0.37\ \ \ \  &\	 \ 0.31\ \ \ \  &\ \	1.16\ \ \ \  \\
Sortino Ratio &\ \ \ 0.22\ \ \ \  &\ \	0.94\ \ \ \  &\ 	0.67\ \ \ \  &\ \ 	0.61\ \ \ \  &\  \	2.97\ \ \ \  \\
\end{tabular}
	\label{tab:tab3}
\end{table}

The option of using a closed/zero exposure strategy means that one could, in practice, amplify the investment/leverage. That is, by doing less active 
daily investment one is saving the action for the right moment. This reduces trading costs as well as reducing exposure  (see Table~\ref{tab:tab5}) yet 
has better profits than the fully exposed methods. However, if one were to scale up {\em pro rata} the investment in Model A, that would in turn increase  profits as a fraction of investment, but at the same time increase its volatility. 

Time is a very important feature of practical investment.  In particular, whether the returns are steady (quasi-stationary) rather than a sequence of troughs and peaks. A stochastic process, as it evolves can touch stop-loss barriers which could lead to termination of the investment.
On the other hand, a growing value of investment  allows reinvestment and compounding of the gains (and losses).  
 
\begin{figure}[t]
	\centering
	\caption{Cumulative Daytime Percentage Profit Over Six Years (Jan 2018 – Dec 2023).}
	\vspace{0.25cm}
        \includegraphics[width=12cm]{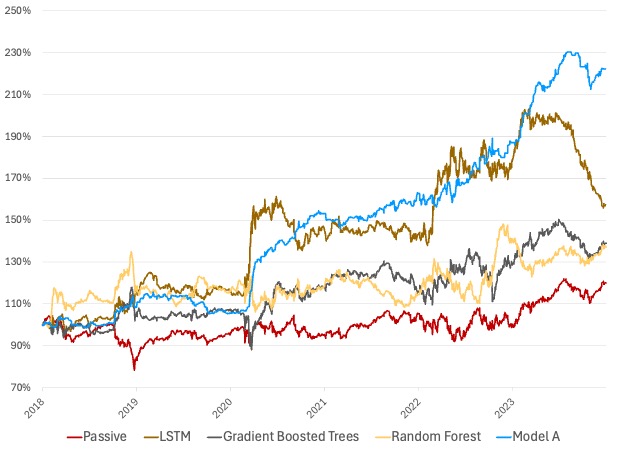}
	\label{fig:fig3}
\end{figure}

We follow the course of the  different investment models over the 6-year test window in Figure~\ref{fig:fig3}. 

It is clear that, while the passive line (Figure~\ref{fig:fig3}) is rather steady (on this vertical scale)  all the tested models have wide variations. While the models outperform the passive benchmark, there are extreme lows and highs, with the exception of Model A, which has reduced fluctuations  over different time periods. We note that this figure uses daily compounding as an illustration, for the passive as well as active investments. In practical terms, futures contracts are discrete and therefore increasing or decreasing exposure has granularity, depending on the number of contracts in play. Since contracts have a fixed price, this in turn depends on the amount under investment. Over this time-frame Model A shows a significant and consistent outperformance compared with the passive, and with lower volatility. The second-best performer is the LSTM but its higher return comes at the expense of higher volatility and much deeper drawdowns (see Table~\ref{tab:tab4}).

\begin{figure}[htbp]
	\centering
	\caption{Cumulative Daytime Percentage Profit Over Three Years (Jan 2021- Dec 2023).}
	\vspace{0.25cm}
        \includegraphics[width=12cm]{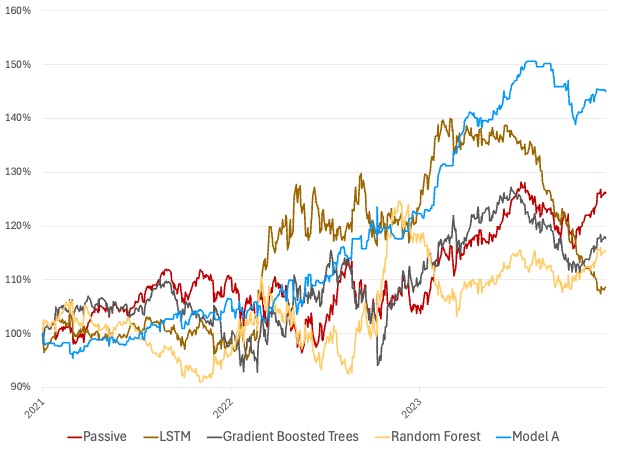}
	\label{fig:fig4}
\end{figure}

In practical terms, futures contracts are discrete and therefore increasing or decreasing exposure has granularity, depending 
on the number of contracts in play. Since contracts have a fixed price, this in turn depends on the amount under investment. 
Figure~\ref{fig:fig4} shows the same results but over a three-year window beginning in January 2021.   In this more recent comparison the 
passive gains (again using compounding) exceed 25\%,  while LSTM has a sharp decrease in profit during 2023. Model A struggles during  early 2021, but returns to profitability. During this shorter time-frame Model A once again outperforms the other models and the passive, with lower volatility. The passive is the second-best performer, doing better than any of the other three models tested.
Additional detailed performance data showing monthly percentage profits for the six years tested is shown in Table~\ref{tab:tab6}.

\begin{table}[htbp]
\small
	\caption{Further Performance Metrics (Based on Monthly Returns Jan 2018 – Dec 2023).}
	\vspace{0.25cm}
	\centering
	\sffamily
	\begin{tabular}{lccccc} 
 &   &   &   { \textsf{\textbf{ Gradient }}}   & { \textsf{\textbf{  Random }}}  &         \\	
  &  { \textsf{\textbf{ Passive}}}    &   { \textsf{\textbf{ LSTM }}}  &  { \textsf{\textbf{Boosted }}} & { \textsf{\textbf{  Forest   }}} &   { \textsf{\textbf{ Model A  }}}      \\	
   &  &   &   { \textsf{\textbf{ Trees }}}  &    &       \\	
      &  &   &    &    &       \\	
Maximum Drawdown &-19.56\% 	&-21.32\%&	-11.77\%	&-12.24\%	&-7.87\%\\
\% Winning Months & 58.33\%&	58.33\%&\ 	51.39\%&\ 	48.61\%	&66.67\%\\
\% Losing Months &  41.67\%	&41.67\%&\ 	48.61\%&	\ 51.39\%&	33.33\%\\
Calmar Ratio &0.70 \ \ &	0.39\ \  &	0.65\ \   &	0.40\ \  	&\  1.88\ \   \\
Information Ratio &-	&1.51\ \ 	&0.83\ \ 	&\ 0.56\ \ \  	&\ \ 2.80\ \ \ \  \\
\end{tabular}
	\label{tab:tab4}
\end{table}

Table~\ref{tab:tab5} illustrates the percentage of daily decisions that are long or short trades and shows the percentage profit contribution. As is noted the passive has 100\% long exposure since it only ever takes a long position. The two classification methods have a larger long bias than their LSTM counterpart. Model A once again stands out, with very low exposure compared to all the other methods. Its total market exposure (long or short) is only 41.95\% compared to 100\% for the others. This indicates the agent-pair tendency to only trade when it has ascertained that there is sufficient likelihood that the trade is worthwhile. Interestingly, Model A is the most accurate both directionally and in terms of profit. It is also necessary to bear in mind that from a fund management perspective, resources not exposed to market risk, can be allocated elsewhere, at minimum in a risk-free investment. This of course would further contribute to increasing the returns generated by Model A, something which is not considered in the results presented.

\begin{figure}[!h]
\caption{Distribution of Daytime Daily Returns (Jan 2018 – Dec 2023).}
\vspace{0.25cm}
\begin{subfigure}{.5\textwidth}
	\centering
	\caption*{ \textsf{\textbf{ (a)   LSTM vs Passive}}}
	\includegraphics[width=0.9\linewidth]{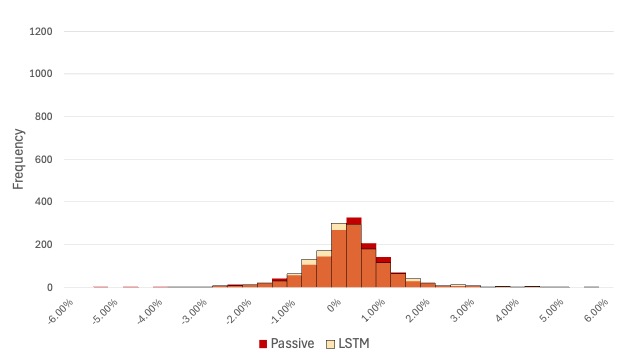}  
\end{subfigure}
\begin{subfigure}{.5\textwidth}
  \centering
  \caption*{ \textsf{\textbf{  (b)    GBT  vs Passive}}}
	\includegraphics[width=0.9\linewidth]{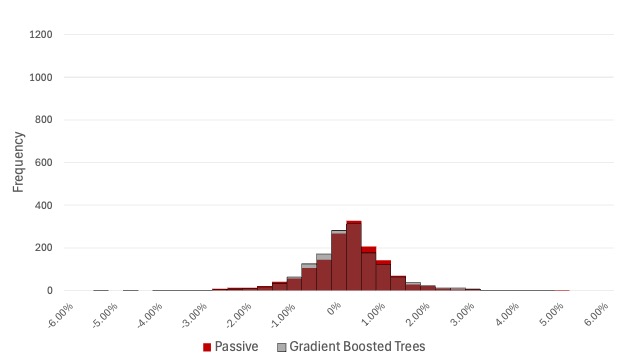}  
\end{subfigure}

\vspace{0.5cm}

\begin{subfigure}{.5\textwidth}
	\centering
	\caption*{ \textsf{\textbf{  (c) Random Forest  vs Passive}}}
	\includegraphics[width=0.9\linewidth]{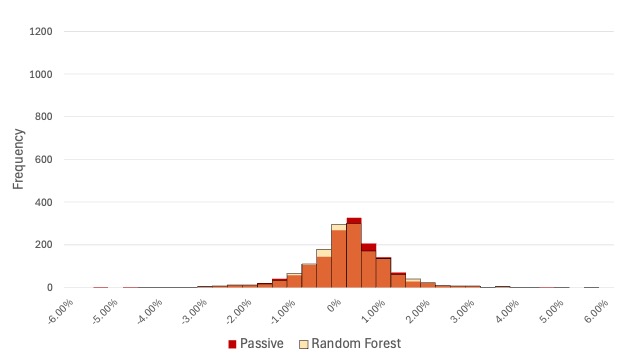}  
\end{subfigure}	
\begin{subfigure}{.5\textwidth}
	\centering
	\caption*{ \textsf{\textbf{   (d)  Model A  vs Passive}}}
	\includegraphics[width=0.9\linewidth]{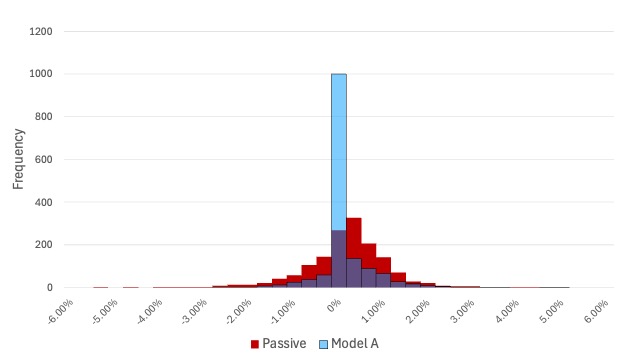}  
\end{subfigure}
\label{fig:fig5}
\end{figure}

The following four charts (Figure~\ref{fig:fig5}) show the frequency distribution of daily daytime returns (\%) (bin size = 0.33\%) for the test sample,  $N=1509 $. 
They compare each of the tested models vs the distribution of the passive benchmark. The accompanying summary statistics are provided in Table~\ref{tab:tab2}. 

\begin{table}[ht]
\small
	\caption{Market Exposure Efficiency and Contribution of Signal Direction to Overall Profit (Based on Daily Daytime Returns Jan 2018 - Dec 2023).}
	\vspace{0.25cm}
	\centering
	\sffamily
	\begin{tabular}{lccccc} 
 &   &   &   { \textsf{\textbf{ Gradient }}}   & { \textsf{\textbf{  Random }}}  &         \\	
  &  { \textsf{\textbf{ Passive}}}    &   { \textsf{\textbf{ LSTM }}}  &  { \textsf{\textbf{ Boosted  }}} & { \textsf{\textbf{  Forest   }}} &   { \textsf{\textbf{ Model A  }}}      \\	
   &  &   &   { \textsf{\textbf{ Trees }} } &    &       \\	
      &  &   &    &    &       \\	
\% Exposure Long    & \ 100.00 \% & 	53.48 \% & 57.19 \%	& 59.77 \%  & 	27.44 \%    \\
\% Exposure Short   &\  \  \  \ 0.00 \%  &  46.52 \%  &  42.81 \%  & 	40.23 \% & 14.51\% \\
Long Contribution   &  100.00  \%  & 	77.31 \%	      &      71.74 \%	    &                    55.10 \%	    &       68.27 \%  \\
Short Contribution  &  \  \  \  \ 0.00 \%	    &     22.69 \%	      &      28.26 \%	      &                  44.90 \%	      &     31.73 \%  \\
\end{tabular}
	\label{tab:tab5}
\end{table}

From Figure~\ref{fig:fig5}, it is very noticeable how different the distribution for Model A is, compared to the other models and the passive benchmark. Its large frequency concentration at 0\% reflects how frequently the agent action makes a decision not to open a position, and thus reduces exposure.

\begin{table}[p]
\small \caption{Monthly Percentage Profits}
	\vspace{0.35cm}
	\centering
	\sffamily
\begin{tabular}{lrrrrrrrrrrrrr} 
\multicolumn{14}{l}{\textsf{\textbf{ Passive}}}    \\
		        & Jan & Feb & Mar & Apr & May & Jun & Jul & Aug & Sep & Oct & Nov & Dec & Year   \\ \\
2023 & 5.78 & 2.22 & 2.05 & 2.67 & -1.19 & 5.97 & 1.44 & -2.06 & -4.43 & -1.41 & 4.21 & 3.54 & 19.83   \\
2022 & -4.12 & 1.81 & 1.06 & -7.49 & 1.95 & -3.44 & 10.56 & -2.41 & -4.30 & 7.36 & 0.44 & -4.53 & -4.49   \\
2021 & -1.13 & 1.29 & 1.20 & 3.43 & -0.05 & 0.26 & 3.58 & 2.80 & -5.35 & 4.22 & -1.66 & 1.56 & 10.18   \\
2020 & 1.67 & -3.31 & 4.09 & 1.90 & -1.34 & -1.66 & 2.28 & 4.04 & -5.70 & -2.56 & -0.65 & 1.45 & -0.32   \\
2019 & 8.63 & 3.16 & -0.21 & 1.30 & 1.31 & 0.43 & 0.24 & -2.58 & -0.50 & 0.16 & 1.97 & 0.67 & 15.16  \\ 
2018 & 3.10 & -3.51 & -4.77 & -3.36 & 3.70 & -0.53 & 3.07 & 2.79 & -0.14 & -8.75 & 1.96 & -10.79 & -17.06 \\ \\
\multicolumn{14}{l}{\textsf{\textbf{ LSTM}}}   \\ 
 & Jan & Feb & Mar & Apr & May & Jun & Jul & Aug & Sep & Oct & Nov & Dec & Year   \\ \\
2023 & 9.11 & 2.12 & 0.07 & 0.76 & -1.25 & 0.64 & -2.71 & -4.45 & -5.09 & -3.10 & -4.56 & -2.98 & -11.60 \\
2022 & -2.72 & 11.14 & 9.50 & 0.83 & 0.12 & 4.22 & -2.90 & 5.53 & -5.08 & -1.84 & 1.05 & 3.44 & 24.32  \\
2021 & 2.37 & 1.15 & -2.66 & -1.60 & -0.03 & 2.88 & 0.81 & -2.24 & -1.09 & 1.72 & -2.54 & 0.23 & -1.21  \\
2020 & -0.51 & 3.13 & 31.13 & -2.09 & 1.12 & 1.25 & -4.18 & -1.84 & -7.29 & 1.52 & 2.71 & 1.45 & 24.39  \\
2019 & 4.36 & 2.72 & -5.33 & 0.30 & -2.13 & 0.91 & 0.13 & -3.12 & 0.11 & 3.66 & 0.05 & -1.10 & 0.14  \\
2018 & -1.25 & 3.07 & -3.84 & 0.65 & 3.70 & -1.75 & 1.26 & 1.19 & 0.11 & 6.07 & -0.92 & 7.61 & 16.43  \\ \\
\multicolumn{14}{l}{\textsf{\textbf{ GBT}} }   \\ 
 & Jan & Feb & Mar & Apr & May & Jun & Jul & Aug & Sep & Oct & Nov & Dec & Year  \\  \\
2023 & 2.74 & 5.95 & -0.47 & 2.34 & 1.36 & 3.63 & -3.20 & -1.37 & -2.79 & -4.46 & 2.23 & 2.57 & 8.29  \\
2022 & -1.40 & 2.93 & 2.67 & 1.94 & 2.82 & 0.48 & 2.54 & -4.24 & -3.59 & -0.65 & 9.47 & -1.88 & 10.87  \\
2021 & 4.94 & 0.26 & 1.15 & 0.04 & -1.40 & -1.12 & 2.48 & 3.03 & -5.35 & -0.10 & -0.87 & -4.53 & -1.92  \\
2020 & -1.35 & -5.11 & -2.39 & 7.02 & 4.72 & 9.28 & -3.41 & -0.31 & -3.14 & 1.87 & 2.36 & 1.34 & 10.28  \\
2019 & -0.75 & -0.78 & 1.22 & -1.21 & 2.52 & -0.63 & -0.97 & 0.21 & 0.39 & 3.85 & -0.18 & -0.54 & 3.04  \\
2018 & 0.77 & -3.51 & 0.03 & 0.75 & 2.30 & -1.28 & -1.19 & -1.28 & 1.14 & 10.89 & -1.03 & -3.04 & 3.89  \\ \\
\multicolumn{14}{l}{\textsf{\textbf{ Random Forest}} }    \\ 
 & Jan & Feb & Mar & Apr & May & Jun & Jul & Aug & Sep & Oct & Nov & Dec & Year \\ \\
2023 & -6.26 & -3.55 & 2.33 & 0.31 & -0.91 & 5.12 & -1.98 & 0.90 & -3.28 & 1.07 & 0.90 & 3.14 & -2.74  \\
2022 & 6.14 & 5.74 & -3.72 & -2.56 & -1.72 & -3.92 & 0.44 & 2.23 & 0.94 & 15.03 & 5.41 & -1.37 & 23.25 \\ 
2021 & 2.26 & 2.63 & -0.46 & -3.36 & 0.61 & -0.29 & -3.11 & -0.76 & -5.42 & 0.09 & 1.63 & 2.65 & -3.82 \\
2020 & 3.49 & -5.98 & -2.54 & 2.42 & -1.14 & 4.55 & -2.15 & 0.08 & -0.72 & 3.28 & -1.26 & 2.66 & 2.17 \\
2019 & -1.09 & -1.68 & -1.78 & 0.04 & -3.11 & 2.94 & 0.10 & 8.87 & -2.05 & -1.54 & -1.95 & 0.11 & -1.67 \\ 
2018 & -0.83 & 12.92 & -2.73 & -1.81 & 4.76 & 2.13 & -0.70 & -2.26 & -0.64 & 12.23 & -0.45 & -3.93 & 18.44 \\ \\
\multicolumn{14}{l}{\textsf{\textbf{ Model A}} }   \\ 
 & Jan & Feb & Mar & Apr & May & Jun & Jul & Aug & Sep & Oct & Nov & Dec & Year \\ \\
2023 & 2.58 & 4.57 & 6.26 & 0.43 & 0.47 & 3.94 & 2.98 & -0.69 & -2.45 & -4.82 & 3.07 & 1.33 & 18.57  \\
2022 & -3.39 & 2.22 & 1.26 & 3.21 & 0.39 & -0.50 & 4.48 & 2.53 & 1.46 & 0.24 & -1.62 & 3.97 & 14.87  \\
2021 & -2.23 & -1.58 & 1.41 & 1.34 & -0.19 & 0.25 & 2.26 & 1.60 & -1.85 & 3.20 & -0.26 & 2.52 & 6.49  \\
2020 & 0.44 & 0.79 & 16.42 & 6.83 & 2.40 & 1.76 & -0.27 & 2.47 & 3.70 & 2.57 & 1.33 & 0.42 & 45.12  \\
2019 & 5.43 & -1.16 & 0.78 & -0.81 & 0.59 & -0.07 & -1.95 & -0.99 & -1.62 & -2.63 & 1.19 & -1.43 & -2.87  \\
2018 & 0.77 & 0.31 & 2.11 & -3.09 & -0.42 & -0.36 & 0.15 & 0.39 & 0.72 & 7.12 & 1.11 & -0.30 & 8.52  \\
\end{tabular}
	\label{tab:tab6}
\end{table}

\section{Conclusions}

The results show that a Dynamic Deep Neural Network approach such as Model A, with very limited data inputs can significantly outperform its passive S\&P 500 benchmark. The approach differed distinctly from the other binary classifiers in this test, in that Model A chose to decide 
{\em not} to trade more often than it chose to trade. This circumspectness is difficult to engineer in the other classification methods due to the dangers of over fitting. This is a problem that demands to be explored through further research. 
Our findings, using our proprietary Dynamic Deep Neural Network methodology with limited data inputs, showed strong results for short-term directional predictions on US equity index futures. It suggests that Dynamic Deep Neural Networks using ‘small data’ have the potential to be a very promising direction and exciting field of exploration for fund managers interested in deploying innovative AI Trade Decision-making approaches. 
Given Model A’s lack of market exposure, only 41.95\%, it makes its outperformance of the passive strategy and the other models even more impressive. The AI system achieved an annualized return that was $\times $ 4.82 greater than the passive’s annualized return and a profit accuracy rate that was 17.5\% better than the passive’s rate. This illustrates the strength of the model’s ability to identify and avoid loss-making trades. 
Another result that is also worth registering is how uncorrelated all of the test models were in comparison to the passive benchmark.
Finally, it should be noted that Model A is not just a ‘lab-model’ as it is currently deployed as part of Plotinus Asset Management’s actively trading Non-Correlated Alpha Controlled Risk Strategy (this strategy also trades the futures market overnight). Its daytime performance exhibits equivalence to Model A’s results.

\newpage

\bibliographystyle{unsrt}

\newpage 

\section*{Disclaimer}

This Document contains information that may be relevant to Investment Professionals, as such
 it is necessary for the Investment Professional to consider all the risks associated with investing and whether or not a particular investment is appropriate for their requirements and that they are in a position, having the resources necessary to bear any losses consequent of an investment. The value of an investment and any income derived from it may increase as well as decrease and thus the value of an original investment may increase or decline accordingly.
It is expected that an Investment Professional will be aware of the speculative nature of Alternative Investments and the associated risk which they could be exposed to over and above Market Risk. It is expected that they will make themselves aware of the following including, but not limited to, Leverage Risk, Liquidity Risk, Counterparty Risk and Currency Risk.
This presentation does not constitute an offer to invest nor Investment Advice of any kind, it is for informational purposes only and its contents are not specific to any individual recipient.
Plotinus Asset Management does not accept any responsibility or liability for any investment decisions made on reliance of the information contained in this presentation.
The information in this presentation has been prepared on behalf of Plotinus Asset Management and has not been independently audited or verified. The source of the data contained in this presentation is Plotinus Asset Management unless otherwise stated. Plotinus Asset Management does not vouch for nor assume any responsibility for the accuracy or completeness of third-party data.
This presentation is subject to updating and revision at any time and Plotinus Asset Management has no obligation to inform the recipient of any changes made to it. Any opinions expressed in this presentation are solely those of its authors and are subject to change. The information contained in this presentation is considered current up to and including the date printed on the cover page.

\section*{Simulated (Hypothetical) Performance Data }

{\small 
This presentation assumes that all profits were reinvested and that the returns were compounded monthly. PLEASE NOTE: "HYPOTHETICAL PERFORMANCE RESULTS HAVE MANY INHERENT LIMITATIONS, SOME OF WHICH ARE DESCRIBED BELOW.
NO REPRESENTATION IS BEING MADE THAT ANY ACCOUNT WILL OR IS LIKELY TO ACHIEVE PROFITS OR LOSSES SIMILAR TO THOSE SHOWN. IN FACT, THERE ARE FREQUENTLY SHARP DIFFERENCES BETWEEN HYPOTHETICAL PERFORMANCE RESULTS AND THE ACTUAL RESULTS SUBSEQUENTLY ACHIEVED BY ANY PARTICULAR TRADING PROGRAM. ONE OF THE LIMITATIONS OF HYPOTHETICAL PERFORMANCE RESULTS IS THAT THEY ARE GENERALLY PREPARED WITH THE BENEFIT OF HINDSIGHT. IN ADDITION, HYPOTHETICAL TRADING DOES NOT INVOLVE FINANCIAL RISK, AND NO HYPOTHETICAL TRADING RECORD CAN COMPLETELY ACCOUNT FOR THE IMPACT OF FINANCIAL RISK IN ACTUAL TRADING. FOR EXAMPLE, THE ABILITY TO WITHSTAND LOSSES OR TO ADHERE TO A PARTICULAR TRADING PROGRAM IN SPITE OF TRADING LOSSES ARE MATERIAL POINTS WHICH CAN ALSO ADVERSELY AFFECT ACTUAL TRADING RESULTS. THERE ARE NUMEROUS OTHER FACTORS RELATED TO THE MARKETS IN GENERAL OR TO THE IMPLEMENTATION OF ANY SPECIFIC TRADING PROGRAM WHICH CANNOT BE FULLY ACCOUNTED FOR IN THE PREPARATION OF HYPOTHETICAL PERFORMANCE RESULTS AND ALL OF WHICH CAN ADVERSELY AFFECT ACTUAL TRADING RESULTS. 
FUTURES TRADING INVOLVES RISK. THERE IS A RISK OF LOSS IN FUTURES TRADING. }

\section*{Commodity Futures Trading}

Futures trading has large potential rewards, but also large potential risk. You must be aware of the risks and be willing to accept them in order to invest in the futures markets. Do not trade with money you can't afford to lose.
This is neither a solicitation nor an offer to Buy/Sell futures. No representation is being made that any account will or is likely to achieve profits or losses similar to those discussed in this presentation. The past performance of any trading system or methodology is not necessarily indicative of future results.

\end{document}